\documentstyle[12pt,epsfig]{article}
\if@twoside m
\else
\fi

\def\real{{\rm I\kern-.2em R}}
\def\complex{\kern.1em{\raise.47ex\hbox{
	    $\scriptscriptstyle |$}}\kern-.40em{\rm C}}
\def\integer{{\rm Z\kern-.32em Z}}

\title{ Chiral boundary conditions for Quantum Hall systems.}
\author{E. Akkermans and R. Narevich \\ Department of Physics, Technion
, 32000 Haifa, Israel}

\begin{document}
\maketitle
\begin{abstract}
A quantum mesoscopic billiard can be viewed as a bounded electronic 
system due to some external confining potential. Since, in general,
 we do not have access 
to the exact expression of this potential, it is usually replaced
 by a set of boundary conditions.
We discuss, in addition to the standard Dirichlet choice, the other 
possibilities of boundary conditions which might correspond to more complicated 
physical situations including the effects of many body interactions or of 
 a strong magnetic field. The latter case is examined more in details using
 a new kind of chiral boundary conditions for which it is shown that in the
Quantum Hall regime, bulk and edge characteristics can be described 
in a unified way. 

\end{abstract}
\newpage

{\bf 1) Introduction.}

One of the main issues in Quantum Mesoscopic Physics is to study the behaviour
 of many particle quantum systems in confined geometries. For many purposes
the many-body interactions are negligible 
and the problem reduces to those of one particle in a confined geometry.
 The corresponding
 hamiltonian is a sum of a kinetic term and a one body operator describing 
either the confining potential or disorder in the bulk of the system. 
The expression "Quantum Mesoscopic Billiards" (QMB) was coined to describe
 generically this class of problems. The role played by the boundaries in
 the behaviour of QMB is central. In the absence of bulk disorder,
the shape of the boundary determines the nature of the energy spectrum, i.e.
whether or not the system will show quantum signatures of chaos.

The aim of this article is twofold. First, it is to discuss in general
 terms what motivates the choice of a given set of boundary conditions and 
to see under which conditions this choice is justified 
for confined quantum systems in situations other than the QMB
 defined above, for instance for a many-body
 system when the hamiltonian is not anymore quadratic or in the presence of
 a high magnetic field i.e. in the Quantum Hall regime.

{\bf 2) How to choose boundary conditions ?}

Consider the case of a QMB without bulk disorder. It is described by the 
Hamiltonian
$H = - {{\hbar^2} \over {2m}}  \Delta + V(r)$ 
where $V(r)$ is a confining potential. It is built up microscopically from the 
electrostatic description of two electron gases of different dielectric 
characteristics. For a given ratio of the dielectric constants, the effective
 image force is strong enough to keep the electrons localized in a given 
area (the billiard). To know exactly the shape of the potential $V(r)$ and to 
solve for it the Shr\"odinger equation is a hopeless task. Then, under the 
assumption that $V(r)$ has bound states, it is possible to replace this problem
 by a simpler one, supposedly equivalent, defined by 
$H = - {{\hbar^2} \over {2m}} \Delta$ 
and $\psi {|_{\cal B}} =0$ for the wavefunction,
 where the boundary $\cal B$ is obtained from the 
symmetry and the shape of $V(r)$. This is the so called Dirichlet choice and 
it is widely used to describe QMB. A more technical remark is 
perhaps appropriate at this stage. This kind of "box quantization" obtained
 using Dirichlet boundary conditions is also widely used to describe other
 physical situations like, for instance, transport in 
a quantum system. Here, unlike the QMB case, the coupling to the external 
world through "leads" plays a central role and the spectrum of the whole system
 is continuous. This leads very often to ill defined or diverging quantities
 which are regularized using instead a discrete spectrum. To this aim, 
the Dirichlet choice is also used among others assuming that it describes 
hopefully the same physics in the limit when the boundary recedes to infinity.
 I shall not discuss this issue any further (Akkermans 1997).

Although the Dirichlet choice is the most popular for the reasons discussed
 above, it is not the only one and may even lead to unpleasant surprises.
 Consider for instance the case of a confined Dirac particle (a Dirac billiard)
described by:
$$ \left( \begin{array}{ccc} {0} & &  {D^\dagger} \\
{D}& &   {0} \end{array} \right) \left( \begin{array}{c} 
 {u}\\{v} \end{array} \right) = E \left( \begin{array}{c} 
 {u}\\{v} \end{array} \right) $$
instead of a Schr\"odinger hamiltonian. Here, $D$ and $D^\dagger$ are first
 order differential operators (the roots of the Laplacian) and the 
wavefunction $\psi =\left( \begin{array}{c} 
 {u}\\{v} \end{array} \right)$ is a two-component spinor. By demanding Dirichlet
boundary conditions, the problem is overdetermined and $\psi$ is 
identically zero not only on the boundary 
but in the whole system. It is also known for 
this problem that 
other choices of local boundary conditions (e.g. Neumann) lead 
 to difficulties associated with the creation of particle-hole pairs 
(Klein paradox) (Berry, Mondragon 1987).
 This problem is not only an academic curiosity, 
but might be
 relevant if one wants to describe mesoscopic superconducting billiards 
where the spectrum is obtained from the Bogoliubov-de Gennes hamiltonian 
which, when linearized, belongs to the class of Dirac problems.

{\bf 3) Beyond one particle: effective hamiltonians.}

So far we did consider the case of quadratic hamiltonians i.e. the laplacian
 plus a (one body) confining potential. When many-body effects cannot be 
neglected anymore, the situation is far more complicated. A standard 
form for the (tight binding) hamiltonian is 
$$H= {\sum_i} {\epsilon_i}{c_i}^{\dagger}{c_i} +
{1 \over 2} {\sum_{ijkl}}\langle ik|V|jl\rangle {c_i}^{\dagger}{c_k}^{\dagger}
{c_l}{c_j}.$$
The kinetic part is still given (in a second quantized form) by a sum of 
laplacian operators, but the second part associated with the interaction 
is a quartic term. Except for some special cases we do not know how to 
diagonalize such hamiltonians no matter wether the system is bounded or not.
The main issue underlying the search of various approximations is
 precisely to define instead an effective quadratic hamiltonian whose
 parameters depend on the approximation. The well known perturbative or
 variational methods (Hartree Fock, RPA, Bogoliubov...) do fulfill this 
objective. When dealing with confined many-body systems, we need to build 
an effective quadratic hamiltonian whose potential takes into account both
 the many-body effects of the confined electrons but also, just like before, 
the effects of the electrostatic potentials resulting from the interactions 
with
the surrounding environment.


Our choice of boundary conditions for the effective one body (quadratic) 
hamiltonian is now broader and depends on the nature of the confining 
potential.
If it is due to image forces as for the QMB case, then the Dirichlet choice 
will be again justified. But if the confinement is dominated by the many-body
 effects in the system itself, then we might be led to other choices of 
boundary conditions.

For the benefit of the more pragmatically inclined reader, let us illustrate 
these ideas by the example of the Feynman ansatz for N strongly interacting 
bosons (Feynman 1954). The many-body hamiltonian is
$$H = {E_0}- {{\hbar^2} \over {2m}} {\sum_i}{\Delta_i} + V,$$
where $V= {\sum_{ij}} V(|{r_i}-{r_j}|)$ is the interaction potential and 
$E_0$ the ground state energy. The N bosons wavefunctions
 describing the excited
 states is assumed (Feynman ansatz) to be of the form 
$\Psi ({r_1},...,{r_N})= F {\Psi_0}({r_1},...,{r_N})$, where $F= {\sum_i^N}
f({r_i})$ and $\Psi_0$ is the exact (but unknown) ground state wavefunction.
 This form is exact for the non interacting case, but it assumes for 
the interacting one that the interactions build up 
separately (under an adiabatic switching)
 in $F$ and in $\Psi_0$. This approximation may be shown to be 
equivalent (under certain conditions) to the RPA, the generator coordinate
 method (Jancovici and Shiff 1964) or the quasi boson approximation. 
The equation of motion of the complex
 function $f(r)$ (it is not the wavefunction) is obtained by minimizing the
 energy
 $E = {{\langle \Psi | H|\Psi \rangle} \over {\langle\Psi|\Psi\rangle}}$.
 Under the assumption of an incompressible ground state of density $\rho_0$, 
$\delta E=0$ implies
$$-{{\hbar^2} \over {2m}} {\rho_0}{\nabla^2} f = E \int d
\vec{r'} f(r') \rho (r-r'),$$
where $\rho (r-r')$ is the density correlation function in the ground state 
$\Psi_0$. The effective energy $E$ is now given by the quadratic form
$$E =- {\rho_0}{{\hbar^2} \over {2m}} \int d\vec{r} f^*
{\nabla^2}f $$
and to obtain the spectrum, we have to impose boundary conditions on the 
function $f$. Assuming translational invariance, Feynman obtained the well
 known relation $E= {{\hbar^2} \over {2m}} {{k^2} \over {S(k)}}$, where $S(k)$
 is the structure factor. This gives the one branch phonon spectrum for small 
$k$. For a bounded system, relating $f(r)$ to the order parameter, we obtain 
that the fluid velocity  is $\vec{v}(r)= {1 \over m} \vec{\nabla}f$ so that 
the natural boundary conditions are Neumann, $\hat{n}\cdot\vec{\nabla}
f{|_{\cal{B}}}=0$ where $\hat{n}$ is a unit vector normal to the boundary.
The same kind of approach applies to the case of bounded superconductors where 
the natural boundary conditions for the effective quadratic hamiltonian are 
now generalized (de Gennes 1966) to
$$\hat{n}\cdot(-i \hbar \vec{\nabla} - {{2e} \over c}\vec{A}) {\big|_{\cal{B}}}
= i \lambda f$$
where $\lambda$ is finite for the boundary between a superconductor
 and a normal
metal while it is zero for an insulator.

To conclude, it looks to be quite a general result that where the Dirichlet 
boundary conditions are more appropriate for the case of a QMB (i.e. usual 
quantum mechanics), the Neumann (or elastic) boundary conditions appear
 to be the
 natural choice for collective (bosonic)
 excitations (phonons,plasmons...) which do
 appear in the effective quadratic approximations of many-body hamiltonians.  
This is intimately related to the 
 semi-classical nature of these approximations.
 They enable us, starting with the microscopic
description, to reduce the problem to the study of large-scale modes 
for which boundary conditions should be 
formulated, according to macroscopic principles (like continuity of the 
current). This leads usually to Neumann boundary conditions.

{\bf 4) Bounded Quantum Hall systems.}

The remaining part of this article is devoted to the application of the
 previous general remarks to the specific case of bounded electrons in a 
 strong magnetic field i.e. in the Quantum Hall regime. I shall focus 
on the simpler case of non interacting electrons.

The various descriptions of the QHE's developed so far belong to two
 main categories. One is based on a bulk description, i.e. on the properties 
of a Landau like spectrum whose main characteristics are the large degeneracy 
of the ground state (proportional to the surface of the system) and its 
incompressibility, i.e. the existence of a gap between it and the first 
excited state. 
These conditions are enough to observe the quantization of the Hall conductance
(MacDonald 1995). The surprising stability of these properties with 
respect to both disorder and interactions are partly responsible
 for the richness
 of this problem. Various points of view were developed in order to prove
 the quantization of the Hall conductance and among them a successful and 
promising topological approach (Thouless et al. 1982, Avron, Seiler and Simon 
1983). There, using periodic boundary conditions, 
the system has the topology of a torus so that edge physics does not play 
any role.

A second line of thought emphasizes the central role played by the edges. 
It is based on the idea that a magnetic field dependent incompressibility 
always leads to gapless edge excitations. Then, the total current being 
zero in the bulk (but not the current density), the currents in a Hall 
experiment flow along the edges (MacDonald 1995, B\"uttiker 1988).

More recently, these edge states were presented as a possible realization 
of a quasi-one dimensional chiral electron gas. Various phenomenological 
models were developed to describe it, including a chiral Luttinger liquid 
(Wen 1990, Stone 1991).
A global description which would relate these two approaches 
 would be welcome. A microscopic way based on first principles to handle 
this question is difficult. To know the exact spectrum of the system, 
we first need to solve a classical electrodynamic problem to obtain the 
confining potential between two electron gases of different
 dielectric functions
 in a strong and inhomogeneous magnetic field. In the absence of applied
 magnetic 
field, the bulk excitations are plasmons with a dispersion $\omega 
\propto \sqrt{k}$. In the presence of the magnetic field the bulk spectrum 
acquires a gap (Kohn's theorem) equal to the cyclotron frequency and chiral 
edge magnetoplasmons propagating along the boundary do appear with a linear 
dispersion. Various descriptions were proposed to study these edge excitations 
using different density profiles (Volkov and Mikhailov
 1988, Aleiner and Glazman 1994).
 Although these 
approaches do provide a qualitative description of the experimental results 
(Ernst et al. 1996)
 they do not take into account quantum effects related to the
 quantization of the Hall conductance, a point which seems to be important 
experimentally. (Ernst et al. 1996).

It would be interesting to know if the microscopic confining potential
 could be 
replaced by an appropriate choice of boundary conditions which contain the 
same physics. To go further, we first consider the case of an effective one 
particle hamiltonian of the form
$$H = - {{\hbar^2} \over {2m}}(\vec{\nabla} - {{ie}\over {\hbar c}}\vec{A}
{)^2} +V(r,B),$$
where $\vec{B} = \vec{\nabla} \times \vec{A}$ is the inhomogeneous
 magnetic field and $V(r,B)$ the effective confining potential, solution 
of the microscopic electrodynamic problem. To replace $V(r,B)$ by a set of 
boundary conditions, we have two main possibilities. The first one is to 
assume that it results from the electrostatics interactions and depends 
very little on the external magnetic field. This situation is similar to the 
QMB we discussed earlier and then we shall choose Dirichlet
 boundary conditions 
$\psi{|_{\cal{B}}}=0$. If on the other hand, the confining nature of the 
magnetic
 field  plays a role, which is expected at high magnetic fields, then the
 Dirichlet choice might be non correct in the sense that although it 
confines the electrons, it will not be able to reproduce the edge excitations.

We are therefore looking for boundary conditions which connect together the
 bulk and edge properties of a confined Quantum Hall system. In other words,
is there for this problem a generalized Poisson principle for which like in
 electrostatics, the bulk and edge excitations are a consequence one of 
the other ?

{\bf 4.1) Effective boundary conditions.}

To go further, we consider the problem of an electron moving in a magnetic 
field $\vec{B}(r) = B \Theta (r-R)\hat{z}$, i.e. uniform in a
 disc of radius $R$
 and vanishing outside. The hamiltonian can then be written
$$ {{2m} \over {\hbar^2}} H= D D^\dagger -b = { D^\dagger} D +b,$$
where $D = {e^{i\theta}}({\partial_r}+{i \over r} {\partial_\theta} + {{br} 
\over 2})$ and $b= {{eB} \over {\hbar c}}= {1 \over {{l_c}^2}}$ ($l_c$ is the 
magnetic length).

{\bf 4.1.1) The Dirichlet choice.}

Demanding $\psi(R,\theta)=0$ for the wavefunction, 
we obtain the spectrum of Fig.1, as a function of 
the angular momentum $m \in\integer$. The main characteristics of this spectrum
 are 
the following:

1) The lowest Landau level is always below the ground state ($m=0$), although

 exponentially close.

2) For any finite $R$, the ground state is non degenerate.

3) The bulk currents $I = {\sum_{m \leq {m_c}}} {{\partial {E_m}} \over 
{\partial m}}$, where $m_c$ corresponds to the Fermi energy,

 are finite and even large.

4) Since $E_m$ are analytic functions
 of $m$ (described
 as a continuous variable),

 there is no natural splitting in this spectrum between bulk and edge states.

\begin{figure}
\begin{center}
  	\leavevmode
         \epsfig{file=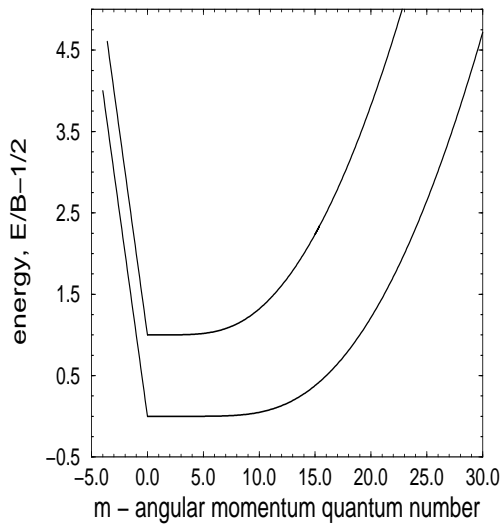,height=7cm,width=8cm,angle=270}
\caption{Spectrum of an electron in the magnetic field - Dirichlet boundary 
conditions.}
\label{fig:1}
\end{center}
\end{figure}

{\bf 4.1.2) The chiral boundary conditions.}

One of the main issues concerning the Dirichlet boundary conditions is that 
they do not provide a sharp dichotomy between bulk and edge states even for 
idealized situations. It is on the other hand a noticeable fact that such a 
dichotomy  naturally exists for a classical bounded system in a magnetic field:
 for a given direction of the field, orbits that lie in the interior of
 the billiard rotate one way, while those hitting the edge make a skipping
 orbit and rotate in the opposite direction. Bulk and edge states are thus 
 distinguished by their chirality relative to the boundary. Recently,
we proposed an 
extension of such a dichotomy to the quantum mechanical case 
(Akkermans, Avron, Narevich and Seiler 1997).
 Consider to that purpose the tangential velocity operator 
in the $m$-sector given by ${v_m}= -{m \over r} + {{br} \over 2}$ with $m \in
\integer$ and consider its spectrum ${v_m}(R)$ projected on the boundary 
$r=R$. The eigenvalues are given by ${v_m}(R) = -{1 \over R}(m - \phi)$, 
where $\phi= { \Phi \over {\Phi_0}}$ is the total magnetic flux through
 the disc 
in units of the flux quantum ${\Phi_0}= {{hc} \over e}$. The chiral boundary 
conditions require:
\begin{eqnarray} D_m{\psi_m}(r)\Big|_{r=R}= 0,
\ &{\rm if}&
\ v_m(R)= -{1 \over R}(m - \phi) > 0;\nonumber\\
{\partial_r}{\psi_m}(r)\Big\vert_{r=R}=
0, \ &{\rm if}&\
v_m(R)= -{1 \over R}(m - \phi)\leq 0,
\label{spn}
\end{eqnarray}
where ${D_m}= ({\partial_r} + {v_m})$.
 The first condition, as a generalization of the classical
 case will correspond to a bulk electron for which we demand elastic 
boundary conditions (${D_m}{\psi_m}(R) =0$), while the second condition will 
describe an edge electron for which we demand Neumann boundary conditions.
 These
 non local (spectral) boundary conditions are relatives of the boundary
 conditions introduced by Atiyah, Patodi and Singer (APS) in their study of
 Index theorems for Dirac operators with boundaries precisely for the reasons 
we discussed earlier(Atiyah, Patodi and Singer 1973). However, they did 
choose for edge states
 Dirichlet instead of Neumann 
boundary conditions here considered for a reason I shall 
discuss later on. It can be checked directly that this choice preserves 
gauge invariance and defines a self-adjoint eigenvalue problem.
 The energy spectrum can be described in terms of special functions and is 
shown on Fig.2. The Hilbert space ${\cal H}$ it defines is the direct sum of 
two orthogonal infinite dimensional spaces ${\cal H}_b$ and ${\cal H}_e$ 
corresponding respectively to bulk and edge states. In contrast to the 
Dirichlet case, the ground state of the bulk spectrum corresponds precisely 
to the lowest Landau level and has a degeneracy given by the integer part 
$[\phi]$ of the total magnetic flux through the disk. The first excited 
bulk state is separated from the ground state by a gap equal to the cyclotron 
frequency which ensures incompressibility. Finally, the total current in the 
ground state $I = {\sum_{m \leq [\phi]}} {{\partial {E_m}} \over 
{\partial m}}=0$, a property of the lowest Landau level. 

The edge spectrum is, in contrast, gapless in the thermodynamic limit and has 
a linear dispersion for low excitation energies with a "sound velocity"
 proportional to $\sqrt B$. The justification of Neumann boundary 
conditions for the 
edge states instead of the Dirichlet original choice of APS comes from the 
requirement of having continuous energy curves between bulk and edge energies. 
 Moreover with the Dirichlet choice, our edge states would have been
 pushed away
 from the edge. 

\begin{figure}
\begin{center}
  	\leavevmode
	\epsfig{file=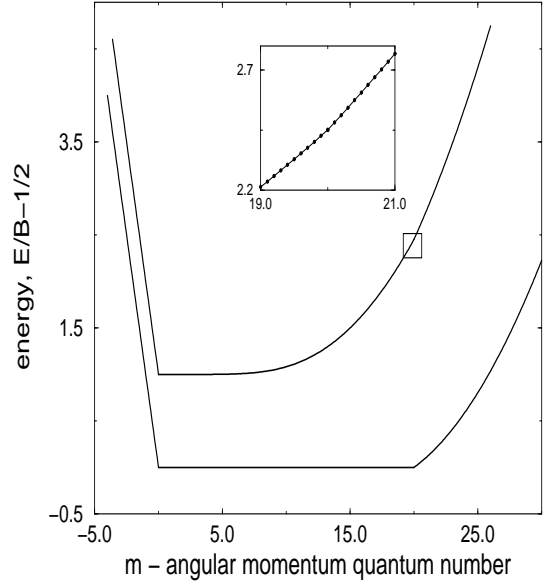,height=7cm,width=8cm,angle=270}
\end{center}
\caption{Spectrum of an electron in the magnetic field - chiral boundary 
conditions (inset - enlarged box,
showing a cusp between bulk and edge states).}
\label{fig:2b}
\end{figure}

{\bf 4.2) Spectral flow.}

Another interesting difference which does appear in our chiral
 boundary conditions is the discontinuity in the first derivative of the
 energy curves. Since this derivative is proportional to the current, such a 
discontinuity would correspond to a non conservation of the electric charge.
 Consider now adding an additional time dependent Aharonov-Bohm magnetic flux  
 at the center of the disc which may fulfill the role of a battery (emf).
Then, by changing it by one unit of
 quantum flux, one state moves from the bulk to 
the edge Hilbert spaces. Then, our boundary conditions allow for counting the 
states that move from bulk to edge. A similar spectral flow takes also place 
in the Dirichlet case but only in a qualitative sense. Here, we can
relate this charge transport (the Hall conductance)
 to a topological index which characterizes 
the spectral flow (Akkermans and Narevich 1997). 

{\bf 4.3) A chiral hamiltonian for the edge states.}

The phenomenological description of the chiral edge states is based on the 
observation that the bulk of the Hall liquid being incompressible and 
irrotational below the Kohn gap, the only low energy excitations are on the 
 boundary and may be represented by chiral bosons derived from a Kac Moody 
algebra (Wen 1990, Stone 1991). The corresponding hamiltonian can be
 derived, but a central 
problem is then to relate it to the bulk quantities. Using our
 boundary conditions, 
this bulk-edge relation naturally comes in.
 Consider the field operators
$\hat{\Psi}(r) = {\sum_{m=0}}^{\infty}{a_m} {\psi_m}(r)$, where ${\psi_m}(r)$ 
are the eigenfunctions of the one particle hamiltonian and $a_m$ the
 annihilation operator of a state of angular momentum $m$. Then, up to a 
constant, the total hamiltonian can be written in a second quantized form as
 $$ H= {\sum_{m=0}}^{\infty}{a_m}^\dagger{a_m} \int {d^2}r {\psi_m}^*(r) 
D\,D^\dagger {\psi_m}(r).$$
Integrating by parts, we obtain
$$ H= {\sum_{m=0}}^{\infty}{a_m}^\dagger{a_m} \big(-\int_{\cal D} {d^2}r 
(D {\psi_m}(r){)^*}(D {\psi_m}(r)) + \int_{\partial\cal{D}}d\theta
{\psi_m}^*(R,\theta)
\hat{v}(R,\theta){\psi_m}(R,\theta) \big )$$
where the first integral is on the disc, while the second is on 
the circle boundary. These two integrals in parenthesis
 define the energy $E_m$.
The ground state corresponds to the full lowest bulk state i.e. for angular 
momentum up to $m=[\phi]$. The lowest excited states are obtained in the limit 
where $m$ approaches $[\phi]$ from above. In that limit, $D {\psi_m}(r)=0$ and
 only the second integral remains in the hamiltonian which can be rewritten 
using the definition of the tangential velocity operator written above
$${H_{edge}}= -i \int_{\partial\cal{D}}d\theta \hat{\Psi}^\dagger(R,\theta)
({\partial_\theta} +i \phi)  \hat{\Psi}(R,\theta)$$
which is the  Dirac hamiltonian density for a one dimensional chiral fermion 
field whose eigenstates are bosonic excitations.

{\bf 5) Conclusions.}

We first notice that, as pointed out before, the bosonic nature of the edge
 excitations (of linear dispersion) is intimately connected with the choice 
of Neumann boundary conditions for these states.

Then, the question arises of the generalization of this approach to include 
interactions (the fractional Hall case) and disorder. Considering the first 
point, it was shown (Akkermans, Avron, Narevich and Seiler 1997)
 that assuming the Laughlin wavefunction 
for the ground state, and a filling fraction $1 \over M$, ($M$ being an odd 
integer), the chiral boundary conditions give precisely the number of states 
that the Laughlin state as a bulk state can accommodate i.e. 
${N \over {BR^2}} = {1 \over M}$.

For non separable problems i.e. including either bulk disorder or
 different shapes, it is still possible to use chiral boundary 
conditions  and to obtain a splitting of the Hilbert space. In 
general, the splitting of the wavefunctions disappears and
 the states will have instead nonzero parts both in ${\cal H}_b$ and in
 ${\cal H}_e$. 

This approach using non local boundary conditions might be a promising way 
to investigate other not unrelated problems like a rotating superfluid 
where the Coriolis force is analogous to the Lorentz force in a magnetic 
field or the Bogoliubov-de Gennes equations for a superconducting billiard.

{\bf Acknowledgment.}

This is a pleasure to ackowledge fruitful and numerous discussions
 with J.E.Avron.
This work is supported in part by a grant from the Israel Academy of
Sciences and by the fund for promotion of Research at the Technion.

{\bf References.}

AKKERMANS, E., AVRON,J.E.,NAREVICH,R. and SEILER,R.,Cond-Mat/9612063,
 preprint 1996

AKKERMANS, E., 1997,J.Math.Phys.{\bf 38},1781.

AKKERMANS, E.,and NAREVICH,R., in preparation. 

ALEINER, I.L., and GLAZMAN, L.I., 1994, Phys. Rev. Lett. {\bf 72}, 2935.

ATIYAH, M.F., PATODI, V.K., and SINGER, I.M., 1973,  Math. Proc. Camb.

Phil. Soc. {\bf 77}, 43.

AVRON, J.E., SEILER, R., SIMON, B., 1983, Phys. Rev. Lett. {\bf 51},51.

BERRY, M.V., and MONDRAGON, R.J.,1987, Proc. R.Soc.Lond. A {\bf 412},53.

BUTTIKER, M.,1988, Phys.Rev.{\bf 38},9375.

de GENNES, P.G.,1966, Superconductivity of Metals and Alloys, Addison and 

Wesley publishing company, inc.

FEYNMAN,R., 1954, Phys.Rev.{\bf 94},262.

ERNST, G., HAUG, R.J.,KUHL, J.,von KLIZING,K., and EBERL, K., 1996,

 Phys. Rev. Lett. {\bf 77},4245.

JANCOVICI, B., and SHIFF,D.H.,1964, Nuclear Physics {\bf 58}, 678.

MacDONALD, A.,(for a review), Les Houches LXI, 
1994, E. Akkermans, G.

 Montambaux, J.~L. Pichard 
and J. Zinn Justin Eds., North Holland 1995.

STONE, M., 1991, Ann.of Physics, {\bf 207},38.

THOULESS, D.J., KOHMOTO, M., NIGHTINGALE, P., and den NIJS, M., 1982,

 Phys. Rev. Lett. {\bf 49},405.

VOLKOV, V.A., MIKHAILOV,S.A., 1988, Sov.Phys.JETP {\bf 67}, 1639.

WEN, X.G., 1990, Phys. Rev. B {\bf 41}, 12838.

\end{document}